\pdfoutput=1

\RequirePackage{lineno}
\documentclass[article,12pt,onecolumn,superscriptaddress]{revtex4}
\usepackage{amssymb,graphicx,color,amsmath,bm,ifthen}

\definecolor{lightblue}{rgb}{0.17,0.39,1}
\definecolor{lightgreen}{rgb}{0.67,0.81,0.08}
\definecolor{lightred}{rgb}{1,0.05,0.52}

\usepackage[bookmarks, colorlinks=true, breaklinks, pdftitle={C. Moir}, pdfauthor={C. Moir}]{hyperref}
\hypersetup{linkcolor=lightblue, citecolor=lightblue, filecolor=black, urlcolor=lightblue}
\newcommand{\BaFeAsP}{{BaFe$_2$(As$_{1-x}$P$_x$)$_2$}}
\newcommand{\YBCO}{YBa$_2$Cu$_3$O$_{6 + \delta}$}
\newcommand{\BaFeCoAs}{{Ba(Fe$_{1-x}$Co$_{x}$)As$_2$}\ }
\newcommand{\LSCO}{La$_{2-x}$Sr$_x$CuO$_{4}$}

\newcommand{\Tc}{$T_{c}$}
\newcommand{\Hsat}{$H_{sat}$}
\newcommand{\CT}{C/\textit{T}}
\newcommand{\sqrtH}{$\sqrt{H}$}

\begin{document}

\title{Multi-band mass enhancement towards critical doping in a pnictide superconductor}

\author{C.M. Moir\footnote{Correspondance should be addressed to C.M. Moir (email: camilla.moir@gmail.com)}}

\affiliation{Florida State University, Tallahassee, FL 32306, USA}
\affiliation{National High Magnetic Field Laboratory, Florida State University, Tallahassee, FL 32310, USA}

\author{Scott C. Riggs}
\affiliation{National High Magnetic Field Laboratory, Florida State University, Tallahassee, FL 32310, USA}

\author{J.A. Galvis}
\affiliation{National High Magnetic Field Laboratory, Florida State University, Tallahassee, FL 32310, USA}

\author{X. Lian}
\affiliation{Florida State University, Tallahassee, FL 32306, USA}
\affiliation{National High Magnetic Field Laboratory, Florida State University, Tallahassee, FL 32310, USA}

\author{P. Giraldo-Gallo}
\affiliation{National High Magnetic Field Laboratory, Florida State University, Tallahassee, FL 32310, USA}

\author{Jiun-Haw Chu}
\affiliation{University of Washington,  Seattle,  WA  98195, USA}

\author{P. Walmsley}
\affiliation{Stanford University,  Stanford, CA 94305, USA}

\author{Ian R. Fisher}
\affiliation{Stanford University,  Stanford, CA 94305, USA}
\affiliation{SLAC National Accelerator Laboratory, Menlo Park, CA 94025, USA}

\author{Arkady Shekhter}
\affiliation{National High Magnetic Field Laboratory, Florida State University, Tallahassee, FL 32310, USA}

\author{G.S. Boebinger}
\affiliation{Florida State University, Tallahassee, FL 32306, USA}
\affiliation{National High Magnetic Field Laboratory, Florida State University, Tallahassee, FL 32310, USA}

\begin{abstract}
\end{abstract}
\date{\today}
\maketitle

\setlength{\parindent}{0.5cm}
\setlength{\parskip}{0.5cm}

{\bf 
\noindent Abstract

Near critical doping, high-temperature superconductors exhibit multiple anomalies associated with enhanced electronic correlations and quantum criticality. Quasiparticle mass enhancement approaching optimal doping has been reported in quantum oscillation measurements in both cuprate \cite{Ramshaw} and pnictide \cite{Walmsley,Shishido, CarringtonReview} superconductors. Although the data are suggestive of enhanced interactions, the microscopic theory of quantum oscillation measurements near a quantum critical point is not yet firmly established \cite{Arkady}. It is therefore desirable to have a direct thermodynamic measurement of quasiparticle mass. Here we report high-magnetic field measurements of heat capacity in the doped pnictide superconductor \BaFeAsP. We observe saturation of the specific heat at high magnetic field in a broad doping range above optimal doping which enables a direct determination of the electronic density of states recovered when superconductivity is suppressed. Our measurements find a strong total mass enhancement in the Fermi pockets that superconduct. This mass enhancement extrapolates to a mass divergence at a critical doping of} $\mathbf{x~=~0.28}$.

\noindent\textbf{Introduction}

A mass divergence at critical doping has been deduced from quantum oscillation measurements at high magnetic fields up to 90 T in the cuprate superconductor \YBCO\ \cite{Ramshaw}, and in the pnictide superconductor, \BaFeAsP\ \cite{Walmsley, Shishido, CarringtonReview}. These measurements, together with measurements of upper critical magnetic field \cite{Grinenko}, elastoresistivity \cite{Fisher}, and magneto-transport  \cite{AnalytisTransport} in \BaFeAsP, as well as elastic moduli \cite{Bohmer} and specific heat studies \cite{Hardy,Masahito} in other doped BaFe$_{2}$As$_{2}$ compounds (Ba122), provide mounting evidence for a quantum critical origin of the phase diagram in high-temperature superconductors. 

In metals, the electronic specific heat measures the total quasiparticle density of states, which is proportional to the sum of quasiparticle masses on all Fermi pockets in quasi-two-dimensional (2D) systems such as Ba122. The enhancement of the quasiparticle mass in Ba122 approaching optimal doping has been previously deduced from the jump in specific heat at the superconducting transition temperature, \Tc. However, this analysis depends on model assumptions that can only be justified in conventional superconductors, in which the relationship between the specific heat jump and \Tc\ is known \cite{Walmsley, Campanini,Chaparro,Welp,Hardy,Bohmer,Masahito,Zaanen}. What has been missing is a direct measurement of the normal state density of states in high-temperature superconductors, from which the sum of quasiparticle masses from all Fermi pockets can be determined. In this study, we utilize high magnetic fields to fully suppress superconductivity and reveal the doping evolution of the electronic density of states in the normal state of Ba122 superconductors in a broad doping range approaching optimal doping. 

\noindent\textbf{Results}

Figure 1a shows the magnetic field dependence of specific heat divided by temperature, \CT, of \BaFeAsP\  for $x~=~0.46$ (\Tc$~=~$19.5 K) at 1.5 K. Magnetic fields up to 35 T, the highest magnetic field available in which the signal-to-noise necessary for these measurements is achievable, were applied along the c-axis of the samples for all measurements. Two striking features are apparent: \sqrtH\ behavior at low magnetic fields, followed by saturation above a field denoted by \Hsat. In a normal metallic state, one expects no field dependence of \CT. Therefore, we interpret the saturation value of \CT\ at fields above \Hsat, (C/$T $)$_{sat}$, as the specific heat of \BaFeAsP\ in the normal state where superconductivity is fully suppressed (See SI) \cite{AnalytisTransport,Putzke}. The \sqrtH\ behavior of \CT\ is characteristic of a line-node in the superconducting gap of \BaFeAsP, which is corroborated by other measurements \cite{Ye,Hashimoto,Malone,Reid,Tantanar,Murphy,Volovik,Kopnin,Hussey}. The slope of the \sqrtH\ behavior increases with increasing temperature (Figure 1b). Note that at finite temperature the measured specific heat in small magnetic fields is larger than the extrapolated \sqrtH\ behavior. Both of these observations are consistent with the phenomenology of nodal superconductivity, which requires a monotonic increase of the coefficient of \sqrtH\ with increasing temperature and \CT\ $\propto$ H at very low field (SI) \cite{Volovik, Kopnin,Hussey}. Importantly, within the phenomenology of nodal superconductivity, low-field deviation from \sqrtH\ behavior must vanish as zero temperature is approached, because it originates from the excitation of quasiparticles across the vanishingly small superconducting gap near the line-nodes \cite{Volovik, Kopnin,Hussey}. 

These two major features of the observed field-behavior of heat capacity suggest a strategy for the direct determination of the electronic heat capacity of correlated superconductors such as Ba122 pnictides. (\CT)$_{sat}$ at finite temperatures corresponds to a total density of states in the normal state,  i.e. the sum of contributions from the quasiparticles on the Fermi surface, phonons, and, all other low-energy excitations in the system. The density of quasiparticle states that is recovered when superconductivity is suppressed is the difference between the normal-state value of \CT,  (\CT)$_{sat}$, and the value of \CT\ extrapolated to zero field, (\CT)$_{extrap}$. This is depicted in Figure 1b, where we extrapolate the \sqrtH\ dependence to zero field and define (\CT)$_{extrap}$ as the value of \CT\ at the intercept. We then define $\gamma_{H}$  = (\CT)$_{sat}$  - (\CT)$_{extrap}$, as the quasiparticle density of states that superconduct, a quantity that we observe to be temperature-independent in every sample (as illustrated in Figure 1b for $x =$ 0.46). This temperature independence is consistent with what one would expect for a metal. As such, $\gamma_{H}$ represents the electronic specific heat recovered by suppressing superconductivity and is the component of \CT\ directly associated with quasiparticles on Fermi pockets that superconduct. 

Having defined $\gamma_{H}$, the measured \CT\ contains two other contributions. The phonon contribution can be indentified by the \CT\ $\sim T^2$ behavior at low temperatures (Figure 1c). However, the data show that the measured \CT\ has a third contribution which is independent of both magnetic field and temperature over the entire measured ranges of fields $\textrm{(0 T}~<~ H~<~\textrm{35 T)}$ and of temperatures $\textrm{(}\sim\textrm{1.5 K}~<~T~<~\textrm{20 K)}$. This ``background" contribution, $\gamma_{bg}$, can be experimentally identified as the zero-temperature intercept of zero-field temperature scans (Figure 1c). 

Using the physical picture discussed in connection with Figure 1 as a blueprint, we now examine the behavior of the electronic specific heat for several chemical compositions in the range $x~=~$0.44 to $x~=~$0.60 (as color-coded in Figure 2a) for which the highest available magnetic field, 35 T, is sufficient to fully suppress superconductivity.  All samples exhibit both \sqrtH\ dependence at low field and saturation behavior at high field (Figure 2b). We can read the values of $\gamma_{H}$ and $\gamma_{bg}$ directly from the panels of Figure 2b and 2c, respectively. Figure 3 shows the main finding of our high-magnetic-field studies, the doping dependence of  $\gamma_{H}$ (red circles) over the range $0.44~\leq~x~\leq~0.60$. These data provide direct thermodynamic evidence for the enhancement of quasiparticle mass approaching optimal doping in overdoped \BaFeAsP.

\noindent\textbf{Discussion}

To present the dramatic doping dependence of the specific heat data in Figure 3b in terms of the equivalent quasiparticle mass (right axis of Figure 3b) we assume 2D (cylinder-shaped) Fermi surfaces, $\gamma~=~1.5~\sum{n_{i}}{m_{i}}$, where the factor 1.5 depends upon the unit cell volume and atomic mass per formula unit (SI). The equivalent mass associated with $\gamma_{H}$ is enhanced by more than a factor of two over our doping range. 

We include in Figure 3b the mass enhancement that was previously reported from quantum oscillation measurements in \BaFeAsP \cite{Walmsley}. It is important to note that this mass is the mass of a single Fermi pocket ($\beta$-pocket, open black squares) which is the only pocket in this doping range with a quantum oscillation frequency sufficiently resolved to yield a mass. Note that the quantum oscillation mass of the $\beta$-pocket increases by about 40\% over our doping range, less than half of the observed enhancement that we report in $\gamma_{H}$. Together, these observations demonstrate that some Fermi pockets must have an even stronger mass enhancement than that reported for the $\beta$-pocket alone \cite{Walmsley} and therefore some pockets couple more strongly to quantum fluctuations than does the $\beta$-pocket. The precise degree to which each pocket's mass is enhanced remains an open question. The $\beta$-pocket is at the $X$ point of the Brillouin zone \cite{Carrington}, which suggests that it might be the pockets at the center of the Brillouin zone, $\gamma$ and $\delta$, that have stronger mass enhancement and therefore couple stronger to quantum fluctuations in the Ba122 high-temperature superconductor. We note that electronic correlations have been argued to be stronger near the zone center in high-temperature superconducting cuprates \cite{Keimer}.

Contrary to the doping dependence of $\gamma_{H}$, the zero-magnetic field, zero-temperature \CT, $\gamma_{bg}$ (Figure 3a, blue circles), \textit{increases} with increasing doping. While we discuss a few possible physical origins of $\gamma_{bg}$, including nonsuperconducting Fermi pockets and non-Fermionic modes \cite{Kivelson, Kogan} in the Supplemental Information, here we will address a more prosaic interpretation involving pair-breaking, perhaps arising from disorder. If $\gamma_{bg}$ arises from pair-breaking, then the observed increase of $\gamma_{bg}$\ with increased doping would indicate dramatically increased pair-breaking at higher values of x. One would expect that same pair-breaking to have a signature in the magnetic field dependent plots of Figure 2b, namely the low field deviations from \sqrtH\ would be expected to persist to higher magnetic fields as $\gamma_{bg}$ increases, i.e. with increasing x. However, the \CT\ data in Figure 2b clearly shows the opposite trend: as doping increases, the field range over which we observe the low field deviation from \sqrtH\ behavior is readily apparent at x = 0.44, but becomes negligible at higher x. We conclude that this observation renders the pair-breaking scenario as unlikely to be the source of $\gamma_{bg}$. Instead, we propose that $\gamma_{bg}$ reflects a density of states not associated with Fermi pockets that superconduct, although the specific physics underlying $\gamma_{bg}$\ component remains unknown (SI). We therefore return our attention to $\gamma_{H}$, the component of the quasiparticle density of states that participates in superconductivity.

In Figure 4, we plot the inverse total mass as determined from $\gamma_{H}$. Similar to doping behavior of quasiparticle mass in \YBCO \cite{Ramshaw}, the inverse mass appears to vanish linearly with doping as we approach a critical doping near optimal doping, $x~=~0.31$. A linear extrapolation of the inverse mass from our measured doping range indicates a mass divergence at a critical doping of $x~=~0.28 \pm 0.015$ near optimum doping, evidencing a critical slowing of dynamic behavior near a quantum critical point that is common to the Ba122 pnictide and the \YBCO\ cuprate high-temperature superconductors. This reinforces a quantum critical origin of superconductivity in this pnictide high-temperature superconductor, whereby the same quantum fluctuations that lead to superconducting pairing are also responsible for mass enhancement \cite{Keimer, Varma, Anderson}. 

Recent theoretical discussions \cite{Zaanen2, Davison, Hayes} have linked the temperature dependence of the anomalous relaxation rate in high-temperature superconductors with the electronic entropy per unit volume---both of which are linear-in-temperature over a broad temperature range in the normal metallic state. Recent high-field magnetoresistance measurements in \LSCO\ cuprates \cite{Paula} and \BaFeAsP\ pnictides \cite{Hayes} reveal linear-in-magnetic-field dependence of resistivity at very high fields, suggesting linear-in-magnetic-field ``planckian dissipation" \cite{Keimer} common to both families of high-temperature superconductors. However, our data in Figs 1 and 2 indicate a nearly magnetic-field-\textit{independent} electronic specific heat above the saturation magnetic field, \Hsat\, that implies a magnetic-field-independent electronic entropy. Our observations of a mass divergence in the vicinity of a critical doping, together with the nearly magnetic-field-independence of the normal state electronic density of states provide an experimental touchstone for other theoretical discussions of quantum criticality in high-temperature superconductors.

\textbf{Acknowledgments}
The work at the National High Magnetic Field Laboratory is supported by National Science Foundation Cooperative Agreement No. DMR-1157490 and the State of Florida. We thank R. Baumbach, J. Betts, T. Carrington, S. Hartnoll, R. McDonald, K. Modic, B. Ramshaw, O. Vafek, and J. Zaanen for discussions. We thank R. Baumbach, K.-W. Chen, and Y. Lai for their assistance measuring the magnetization of the samples. 
 
\textbf{Author Contributions}
 J.-H.C., P.W., and I.R.F. grew and prepared the samples at Stanford University. C.M.M., S.C.R., A.S., and G.S.B. planned the experiment. C.M.M., S.C.R, J.A.G., P.G.G., and X.L. performed the high-field heat capacity measurements at the National High Magnetic Field Laboratory. C.M.M. and A.S. analyzed the data. C.M.M., A.S., and G.S.B. wrote the manuscript with contributions from all authors. 

\textbf{Data Availability}
The data that are provided here and support the conclusions of this study are available from the corresponding author upon reasonable request.

\textbf{Competing Financial Interests}
The authors declare no competing interests.

\hypersetup{linkcolor=black,citecolor=lightblue,filecolor=black,urlcolor=black}

\newpage

\begin{figure}[ht!!!!!!!!]
\centerline{
\includegraphics[width=\columnwidth]{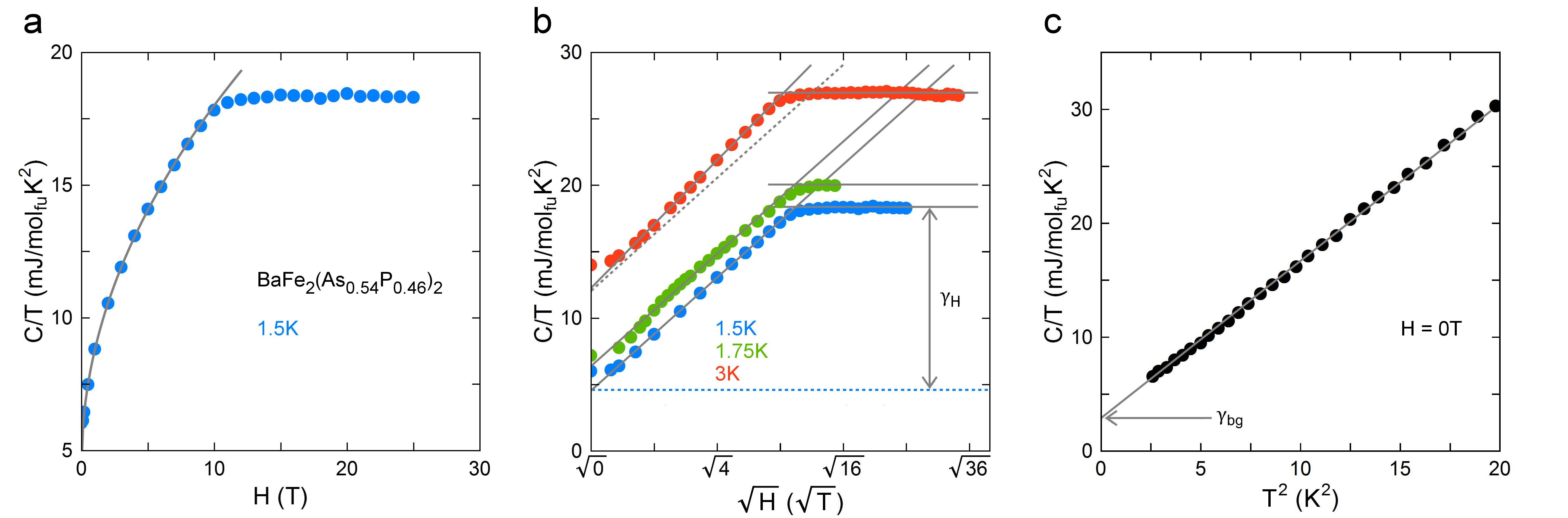}}
\internallinenumbers
\caption{\textsf{Specific heat divided by temperature, \CT, of BaFe$_{2}$(As$_{0.54}$P$_{0.46}$)$_{2}$ (T$_{c}$ = 19.5 K) \textbf{a} Magnetic field dependence of \CT\ at 1.5 K. The gray curve indicates $\sqrt{H}$\ behavior which is consistent with phenomenology associated with a superconducting gap with nodes.\cite{Volovik,Kopnin} \textbf{b} Field dependence of C/\textit{T} plotted against $\sqrt{H}$\ at 1.5 K (blue), 1.75 K (green), and 3 K (red). Solid gray lines indicate the two distinct regimes: $\sqrt{H}$\ behavior and saturation behavior. The slope of the $\sqrt{H}$\ behavior at 1.5 K and 1.75 K  is 4.25 ${\textrm{mJ}}/{\textrm{molK}}^2\sqrt{\textrm{T}}$ and at 3 K is 4.8 ${\textrm{mJ}}/{\textrm{molK}}^2\sqrt{\textrm{T}}$. The dashed, gray line has a slope of 4.25 ${\textrm{mJ}}/{\textrm{molK}}^2\sqrt{\textrm{T}}$ and is provided to compare between the slopes at 1.5 K and 3 K. We define $\gamma_H$ as the difference between the saturation value of C/$T$ and C/$T$ at $H =$ 0 given by the extrapolation of the $\sqrt{H}$\ behavior. \textbf{c} Temperature dependence of C/\textit{T} at zero magnetic field, where the gray line indicates the low temperature specific heat behavior: C/$T = \gamma + \beta T^2$, from which $\gamma_{bg}$ is extrapolated.}}
\label{fig:1}
\end{figure}

\newpage

\begin{figure}[ht!!!!!!!!]
\centerline{
\includegraphics[width=\columnwidth]{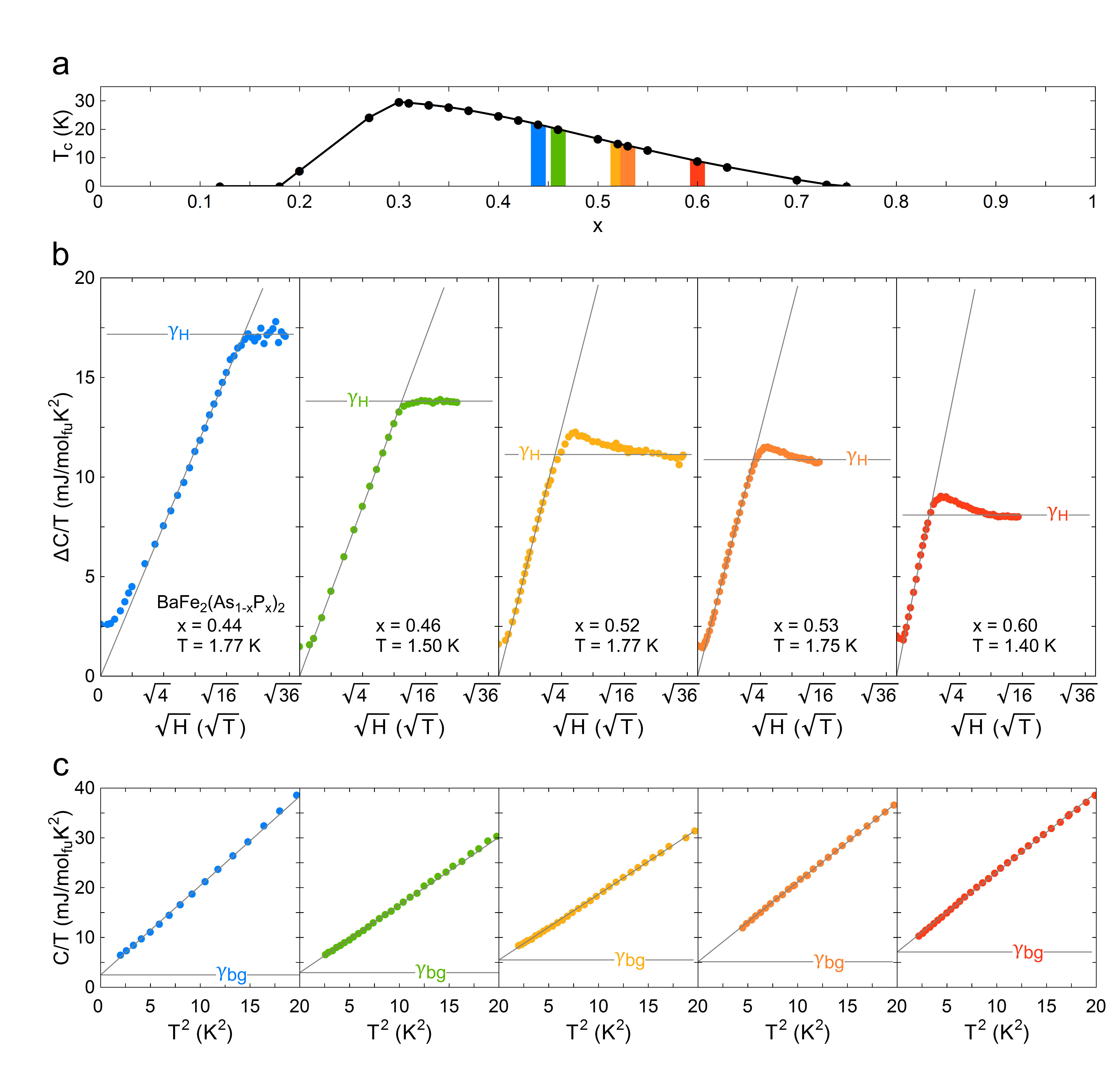}}
\internallinenumbers
\caption{\textsf{\textbf{a} \Tc\ as a function of doping for \BaFeAsP\ aggregated from previous studies \cite{Analytis,AnalytisTransport,Walmsley,FluxGrowth}. Colored lines indicate the doping values of samples studies in this work. \textbf{b} The change in C/\textit{T}, $\Delta\text{C}/T = \text{C}/T(H) - (\text{C}/T)_{extrap}$, from $\gamma_{extrap}$ (see text) of \BaFeAsP\ at low temperatures. Gray lines indicate $\sqrt{H}$\  behavior and saturation at $\gamma_{H}$, which decreases with increasing doping. \textbf{c} Zero field C/\textit{T} as a function of ${T}^2$ in the low temperature regime. Gray lines indicate best agreement to $\gamma + \beta T^2$, the extrapolation of which defines $\gamma_{bg}$.
}}
\label{fig:2}
\end{figure}

\begin{figure}[ht!!!!!!!!]
\centerline{
\includegraphics[width=1\columnwidth]{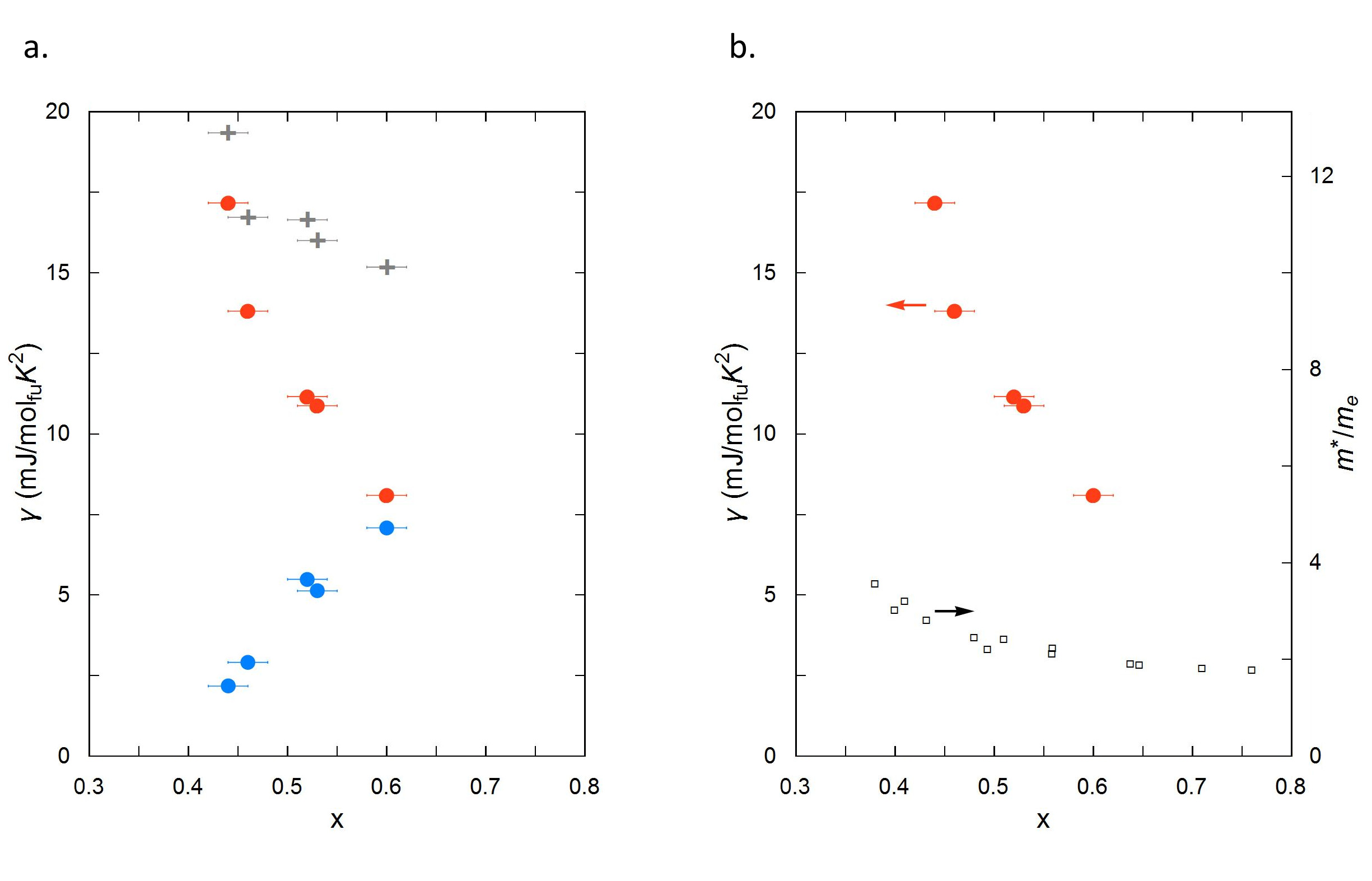}}
\internallinenumbers
\caption{\textsf{\textbf{a} Doping dependence, x,  of the components of the electronic specific heat divided by temperature, $\gamma$, as measured in our study of \BaFeAsP. As phosphorus doping approaches optimal doping (x = 0.31) from the overdoped side, the quasiparticle density of states recovered by suppression of superconductivity, $\gamma_{H}$ (red circles), exhibits an enhancement by more than a factor of two over the doping range studied. The component that persists in the superconducting state in the zero-temperature, zero-magnetic field limit, $\gamma_{bg}$ (blue circles), exhibits the opposite trend with doping, showing a decrease by almost a factor of three over the same range of doping. The sum of $\gamma_{H}$ and $\gamma_{bg}$, (gray crosses) is also plotted and shows an increase by a factor of roughly 1.3. \textbf{b} Doping dependence of $\gamma_{H}$ replotted from panel (\textbf{a}) (red circles) with the corresponding sum of the corresponding of the quasiparticle masses given on the right axis, determined as described in the text. Also plotted is the quasiparticle effective mass of the $\beta$-pocket (empty squares) reported from quantum fluctuation measurements by Wamsley et. al. \cite{Walmsley}. Note that $\gamma_{H}$, the sum of the quasiparticle masses over all pockets taking part in superconductivity, shows a more dramatic enhancement than is seen in the $\beta$-pocket alone.}}
\label{fig:4}
\end{figure}

\begin{figure}[ht!!!!!!!!]
\centerline{
\includegraphics[width=0.8\columnwidth]{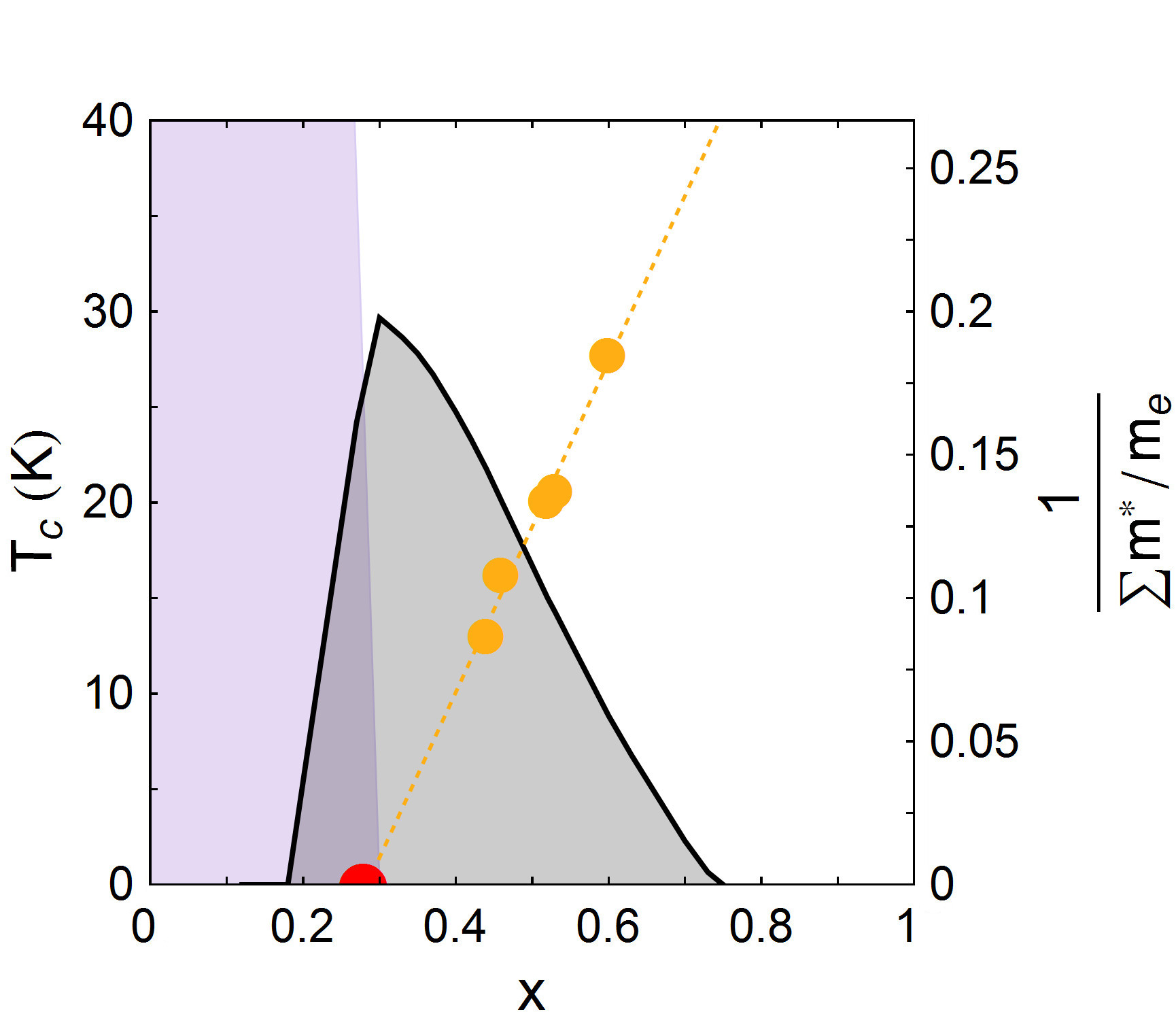}}
\internallinenumbers
\caption{\textsf{Temperature-doping phase diagram of \BaFeAsP. Orange points represent the inverse of the sum of the quasiparticle masses determined from $\gamma_H$. The dashed orange line shows the linear extrapolation of the inverse summed mass to T = 0, the point at which the quasiparticle masses diverge. The black line represents the superconducting transition temperature, $T_c$, aggregated from previous studies \cite{Analytis,AnalytisTransport,Walmsley,FluxGrowth}, and the shaded purple region outlines the spin-density wave regime reported elsewhere \cite{Hashimoto2}.
}}

\label{fig:4}
\end{figure}

\end{document}